\documentclass{PoS}

\usepackage{graphicx}
\usepackage{epsfig}
\usepackage{subfigure}
\usepackage{soul}

\def \aap{A\&A }
\def \apjl{ApJ}

\def \apj{ApJ}

\def \mnras{MNRAS}

\def \nat{Nature\ }
\newcommand{\gray}{$\gamma$-ray~}

\def\msun{{\,M_\odot}}

\title{Galactic centre star formation writ large in gamma-rays}

\ShortTitle{Galactic centre star formation writ large in gamma-rays}

\author{Roland Crocker\thanks{Marie Curie IIF Fellow}\\
        Max-Planck-Institut f{\" u}r Kernphsik, Heidelberg, Germany\\
        E-mail: \email{Roland.Crocker@mpi-hd.mpg.de}}

\abstract{We have modelled the high-energy astrophysics of the inner ~200 pc of the Galaxy with a view to explaining the diffuse, broad-band (radio continuum to TeV $\gamma$-ray), non-thermal signal detected from this region. Our modelling pins down the ISM parameters for the environment wherein cosmic ray (CR) electrons and ions reside in the Galactic centre (GC). We find that the magnetic field in this region is 100-300 $\mu$G, the gas density $\lesssim$ 60 cm$^{-3}$, and that a powerful ($> 200$ km/s) `super'-wind acts to remove $>$ 95\% of the cosmic rays accelerated in the region before they have time to lose their energy in situ. The $\sim10^{39}$ erg/s carried away by the GC cosmic ray protons is precisely enough to energise the $\sim$GeV $\gamma$-ray emission from the Fermi `bubbles' recently found  to extend north and south of the GC out to distances of $\sim$10 kpc, provided that the bubbles constitute thick targets to the GC protons and that the situation has reached steady state. In such a situation of `saturation' the hard, uniform spectrum of the bubbles are explained and secondary electron synchrotron explains the non-thermal microwave emission found in WMAP data mirroring the bubbles. 
Given the very low density of the bubble plasma ($<$0.01 cm$^{-3}$), the $pp$ loss time in the Bubbles is $>$ 5 Gyr. 
Our scenario thus has the startling implication that a GC source of non-thermal particles of time-averaged power $10^{39}$ erg/s has persisted since the youth of the Galaxy.

}

\FullConference{25th Texas Symposium on Relativistic Astrophysics - TEXAS 2010\\
		December 06-10, 2010\\
		Heidelberg, Germany}

\begin{document}

\section{Introduction}

The Galactic centre (GC), besides the intrinsic interest it holds, provides an interesting, potential analogue to the nucleus of a luminous star-burst galaxy.
Indeed, the ISM conditions prevailing in the inner $\sim$ 100 pc (in radius) of the Galaxy -- the region under consideration here -- render it arguably more akin to the environs of  a star-burst than than to the relatively quiescent conditions of the Galactic disk. 
In particular, 
the GC contains something like 5\% of the Galaxy's molecular hydrogen allocation \cite{Morris1996} implying a very high, volumetric-average gas density in the region.
Moreover, the energy-densities of the various GC ISM components are 1--2 orders of magnitude larger than those found locally, as is the areal density of star-formation and attendant supernova activity.
For instance, as we have recently shown \cite{Crocker2010a}, the GC is threaded by a remarkably strong magnetic field of $\sim$100 $\mu$G 
(cf. with $\sim$5 $\mu$G for the Galactic disk).
Such a field 
implies that $\gtrsim$10\% of the Galaxy's magnetic energy is contained 
in only $\lesssim$0.05\% of its volume.

In recent years a picture has begun to emerge that the inner regions  of star-forming galaxies should i) be important sources of $\gamma$-rays in the universe \cite{Pavlidou2002,Thompson2007,Dogiel2009}; ii) drive powerful galactic winds \cite{Dogiel2009} and iii) therefore, be important shapers of the inter-galactic medium, particularly its metallicity \cite{Strickland2009}. 
Interestingly, recent  observations reveal that GC is also a significant \gray source exhibiting both point-like GeV \cite{Chernyakova2011} and TeV \cite{Aharonian2004} emission coincident with Sagittarius A* (at the dynamical centre of the Galaxy and most likely associated with the super-massive black hole found there) and diffuse emission also at both GeV \cite{Digel2009} and TeV \cite{Aharonian2006} energies.

We argue here for a further similarity to star-bursts: 
many direct observations (see Appendix C of \cite{Crocker2010b} and references therein) and, in addition, the non-thermal evidence reviewed below point to the existence of
a power outflow or wind of at least a few 100 km/s out of the GC.
We finally consider the implications of the injection of this  wind fluid -- composed of very hot plasma, cosmic rays and, presumably, `frozen-in' magnetic field lines -- into the Galaxy-at-large and explore a compelling connection between this outflow and the recently-discovered \cite{Su2010,Dobler2009}`Fermi Bubbles'.

\section{Broad-band modelling of the Galactic centre}

We have created a one-zone model of the injection, cooling, and escape of relativistic protons, electrons, and secondary electrons (and positrons) from the inner $\sim$100 pc (in radius) of the Galaxy.
This is the approximate region for which the HESS telescope has reported \cite{Aharonian2006} a diffuse, TeV \gray flux.
To match the TeV  data we have collected archival $\sim$GHz radio continuum data covering the same region (see Appendix D of \cite{Crocker2010c} for radio data sources) and also GeV data \cite{Meurer2009}.
Unfortunately, the latter is heavily polluted by the contribution of i) individual point sources within the field \cite{Cohen-Tanugi2009} including  a GeV source positionally
coincident with Sgr A and  the GC TeV point source \cite{Chernyakova2011}
and ii) by emission from CRs in the line-of-sight along the Galactic plane but out of the GC \cite{Vitale2009} and does not usefully constrain our broadband modelling of the diffuse emission from the region.
As justified elsewhere \cite{Crocker2010c}, in our modelling we assume that the particle astrophysics can be accurately described to be in quasi-steady state 
and that the particle transport timescale is energy-independent.
We also assume homogeneity and isotropy in our modelling.

Our modelling approach is to find -- as a function of environmental and other parameters -- the steady-state populations of relativistic protons and electrons within the region of interest. (Note that in our modelling we neglect for simplicity the poorly-constrained ionic component of the CR hadronic population heavier than protons.) 
We then self-consistently determine (for the same environmental parameters) the radiative output of these populations. 
Relevant radiative processes are neutral meson decay for protons and synchrotron, inverse-Compton and bremsstrahlung for electrons.
We self consistently track both primary and secondary electron emission in our radiative modelling.
Finally, we use a $\chi^2$ minimization procedure to determine the parameters describing the proton and electron populations whose radiative outputs in the
given ISM environment
give the best fit
to the 
$\sim$ GHz radio continuum spectrum and the $\sim$TeV $\gamma$-ray spectrum detected from the HESS field (as previously noted, the GeV data only define upper limits to the broadband emission from the region and we do not attempt to fit them directly).

Environmental parameters that vary within our modelling are magnetic field $B$, ambient hydrogen number density $n_H$ (in whatever form), and the energy-independent timescale over which
particles are advected from the system, $t_{esc} \equiv h/v_{wind}$, where $h \equiv 8$ kpc $\tan(0.3^\circ) \simeq 40$ pc.
(For the reasons explained at length in ref. \cite{Crocker2010c} advection must be the dominant CR transport mechanism in the GC environment so we do not consider CR diffusion in our modelling)

We assume that protons and electrons are injected into the GC ISM with distributions governed by power laws in momentum (with identical spectral indices $\gamma_p = \gamma_e = \gamma$). 
The relative normalization of the {\it injection} distribution of electrons to that of protons (at relativistic energies) is given by the coefficient $\kappa_{ep}$ which is also left as a free parameter in our modelling.
The absolute normalization of the distribution of protons at injection is, finally, specified by $\dot{Q_p}$ (in units eV$^{-1}$cm$^{-3}$ s$^{-1}$), also a free parameter within the model.

\section{Implications of Galactic centre  modelling}

\begin{figure}
\epsfig{file= 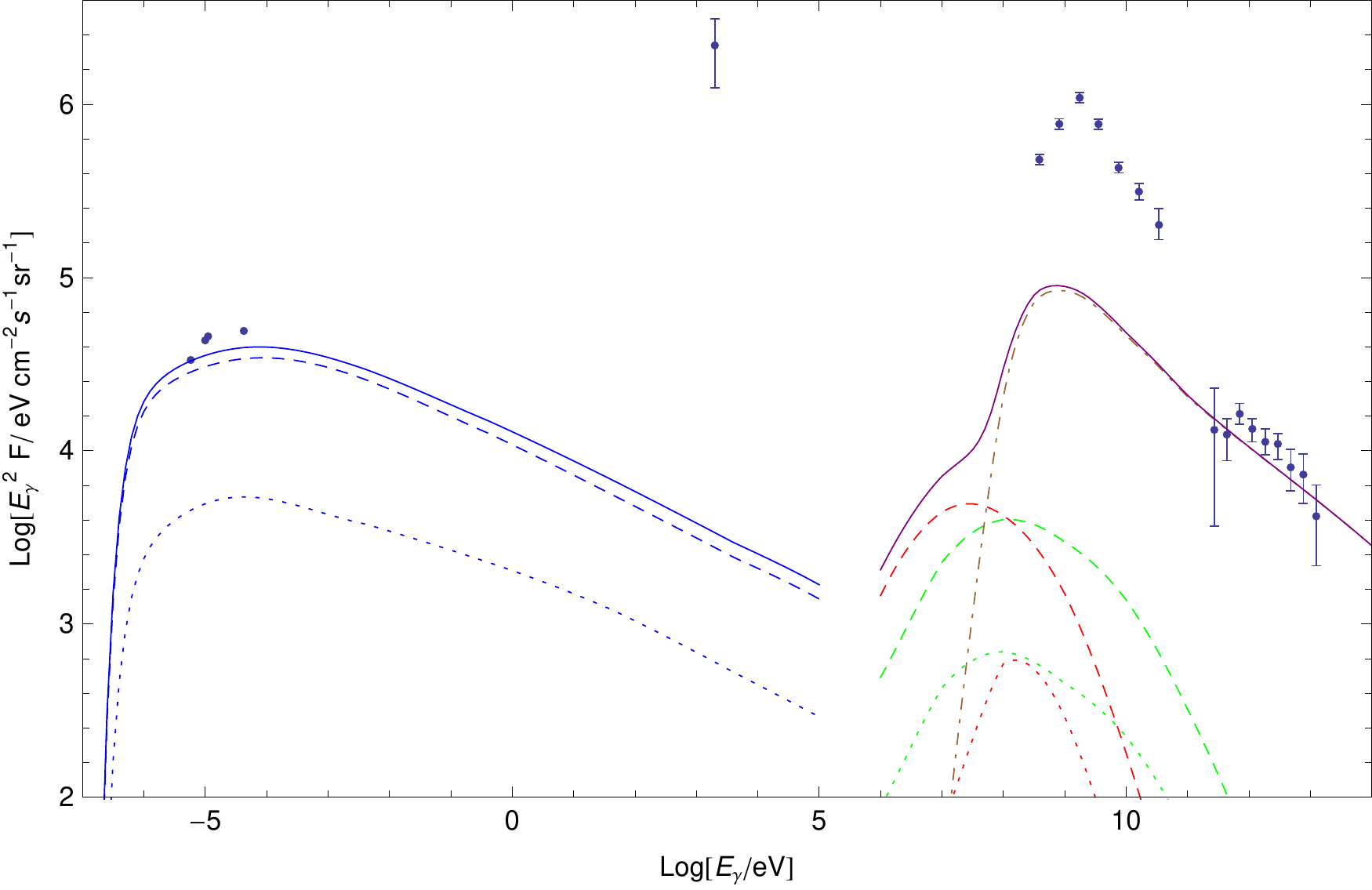,width=0.5 \columnwidth}
\caption{HESS field broadband spectrum energy distribution for the best-fit case. Fitted parameters include magnetic field amplitude ($\sim 200 \ \mu$G), ambient gas density ($6$ cm$^{-3}$), at-injection electron-to-proton ratio at 1 TeV ($\sim$0.005), and wind velocity ($\sim$400 km/s). 
Curves are divided into (i) {\bf dashed} -- primary electron emission; (ii) {\bf dotted} -- secondary electron (and positron) emission; and (iii) {\bf solid} -- total emission at a given photon energy. Emission processes are: {\bf blue} -- synchrotron; {\bf red} -- bremsstrahlung; {\bf green} -- inverse Compton; and {\bf brown, dot-dashed} -- neutral meson decay. The total $\gamma$-ray flux is shown in {\bf purple}. 
Data are from (at $E_\gamma \sim 10^{-5}$) radio, 
(at $E_\gamma \sim$ 2 keV) {\it Ginga} \cite{Yamauchi1990} with ,
(at $E_\gamma \sim$ 1 GeV $\equiv 10^9$ eV) Fermi \cite{Meurer2009}, and (at $E_\gamma \sim$ 1 TeV $\equiv 10^{12}$ eV) HESS observations \cite{Aharonian2006}.
We only display modelled synchrotron emission at radio wavelengths; other processes that combine with this to give the region's {\it total} radio emission are not pictured.
Finally, note that the Fermi data points define only {\it upper limits} to the diffuse emission.
\label{plotBrdBndSpctraBestFit}}
\end{figure}

Our fitting procedure finds acceptable fits for magnetic field amplitudes around
$\sim$ 100 $\mu$G, wind speeds around $v_\textrm{\tiny{wind}} \sim $ few$ \times 100$ km/s, 
total power in all non-thermal particles of $\sim 10^{39}$ erg/s,
gas densities around $n_H \sim 10$ cm$^{-3}$, injection spectral indices $\gamma \sim 2.4$, 
ionization rates of $\zeta \sim 10^{-15}$ s$^{-1}$ and electron to proton ratios (at injection at 1 TeV) of $\kappa_{ep} \sim 10^{-2}$.
Below we illustrate (figs.~\ref{plotBrdBndSpctraBestFit}  the modelled broadband spectrum for the best-fit case 
(which achieves $\chi^2 \simeq 7.9$ for $dof = 9$).

On the basis of our modelling we can determine a number of interesting facts about the GC.
Firstly, the totality of non-thermal signals require a contribution from both primary electrons and protons. 
Neither scenarios where primary electrons alone (in which the observed TeV emission is mostly provided by IC emission) nor where primary protons alone (in which secondary electrons supply the observed synchrotron radiation) provide acceptable fits to the data.
This latter may be in contrast to the case presented by luminous star-bursts where it has been claimed \cite{Thompson2006} that secondary electrons probably do supply most of the observed synchrotron emission.

Secondly the gas environment where the non-thermal radiation is being generated is less dense than $\sim 60$ cm$^{-3}$ (at 2$\sigma$ confidence) with a best fit value close to 1 cm$^{-3}$.
Given that the upper end of the allowed $n_H$ range is less than the volumetric average gas density through the region ($\sim 120 $ cm$^{-3}$) this is an indication that CRs -- even the $>10$ TeV protons responsible for generating the TeV $\gamma$-ray emission do {\it not} penetrate into the densest gas in the region (where star-formation is occurring).
This is, again, apparently in contrast to the situation presented by star-bursts where, it has recently been claimed \cite{Papadopoulos2010}, CRs modify the gas conditions and chemistry where star-formation is occurring, potentially biasing the initial stellar mass function towards more massive stars.

This result -- that GC cosmic rays do not penetrate into the dense gas -- is consistent with another finding: a powerful, star-formation driven superwind blows out of the region with a speed of $\gtrsim 200$ km/s (see \S \ref{sctn_outflow} below).
Given the  wind, comparison of relevant timescales indicates that CRs do not remain long enough in the region to penetrate into the dense molecular gas.

We note that the CRs, however, do apparently constitute important sources of heat and ionization for the warm, diffuse molecular gas phase enveloping the molecular gas cores in the GC environment.

\subsection{Existence of an outflow}
\label{sctn_outflow}

Even without detailed modelling, evidence for a strong outflow can be gathered from a comparison of the thermal and non-thermal signals detected from the inner $\sim 100 $ pc (in radius) of the Galaxy \cite{Crocker2011a}.
At TeV energies, the luminosity expected from the inner $1.6^\circ$ of the Galaxy
were the system to be  calorimetric to all (sufficiently energetic) cosmic rays is given by:
\begin{equation}
L_\textrm{\tiny{TeV}}^\textrm{\tiny{thick}} \simeq \frac{1}{3} \times \eta_\textrm{\tiny{CR}} \times f_\textrm{\tiny{TeV}} \times L_\textrm{\tiny{SN}} \ \simeq 2 \times 10^{37} \ \textrm{erg/s} \ \eta_{0.10} \left(\frac{\nu_{SN}}{0.04/\textrm{century}}\right) \left(\frac{E_{SN}}{10^{51} \ \textrm{erg}}\right)
\label{eqn_hadronic}
\end{equation}
where $\eta_\textrm{\tiny{CR}}$ is the fraction of supernova mechanical power that goes into non-thermal protons normalized to $\eta_{0.10} \equiv \eta_\textrm{\tiny{CR}}/0.1$ and  $f_\textrm{\tiny{TeV}} \simeq 0.05$ is the fraction of total CR proton power in protons sufficiently energetic to generate TeV $\gamma$-rays (adopting  $\sim 10$ TeV for the mean energy of  the parent proton of a TeV $\gamma$-ray and spectral index $\gamma$ = 2.3 in agreement with the TeV spectrum). 
The ground-based  $\gamma$-ray telescope HESS has detected \cite{Aharonian2006} hard spectrum, 
diffuse $\sim$TeV $\gamma$-rays from precisely the same inner $1.5^\circ$ region as noted above. 
This flux, though interpreted originally in the context of an explosive injection of cosmic ray hadrons \cite{Aharonian2006} in a single GC event a few thousand years ago, nevertheless proffers an upper limit on the steady-state $\gamma$-ray intensity from the region.
The total, $>$TeV luminosity of the region is $1.2 \times 10^{35}$ erg/s, which, given the above, is only about 1\% of
the luminosity expected were the system calorimetric for cosmic ray protons.
Thus the vast majority of the power injected into GC
cosmic rays must be carried outside the region.
The same conclusion can be reached from consideration of the non-thermal radio continuum emission from the region \cite{Crocker2011a} and even from the heavily-polluted GeV emission; both these demonstrate that at least 90\% of the power injected into the region's non-thermal proton and electron populations is lost non-radiatively from the region.
The GC, then, loses
$\eta_{CR} \times \nu_{SN}  \times E_{SN} \sim 0.1 \times 0.04/\textrm{century} \times 10^{50} \textrm{erg} \sim 10^{39}$ erg/s in hard-spectrum CRs into the general Galactic environment.

At the heuristic level presented here the conclusion that there is a strong wind out of the region would seem to require that i) we have correctly determined the supernova rate in the region and that, ii) indeed, $10^{50}$ erg per supernova is injected into non-thermal particle populations.
What if our supernova rate is off-beam or, for whatever reason, GC supernova remnants (SNRs) are comparatively inefficient CR accelerators? Would this not, then, imply that we could no longer infer the existence of a wind?
In fact
our detailed modelling generates an {\it independent} constraint on the power injected into CRs in the region.
At $\gtrsim 10^{39}$ erg/s,
 fortuitously or not, this power is essentially what we expect given our supernova rate determination with $10^{50}$ erg per supernova injected into non-thermal particles assumed.
So the conclusion that there is a star-formation-driven wind out of the region seems to be robust. 
What is more, we reach also reach the interesting conclusion that GC SNRs are at least as efficient as the Disk variety as CR accelerations.

\section{Connection to the `Fermi bubbles'}

Recently NASA announced the startling discovery \cite{Dobler2009,Su2010} by $Fermi$ 
of two enormous gamma-ray emission
structures that hang like lightglobes above and below the centre of the
Milky Way. These `Fermi bubbles' extend an
astounding 10 kpc from the plane of the Galaxy. 
At lower Galactic latitudes these structures are coincident with a non-thermal microwave `haze' found  in WMAP 20-60 GHz data  \cite{Finkbeiner2004,Dobler2008} and an extended region of diffuse X-ray emission detected by ROSAT \cite{Snowden1997}.
Thus far
the Bubbles have been typically understood as illuminated by a mysterious
population of youthful and highly energetic ($\sim$TeV) electrons of age 10 million years or so which simultaneously inverse-Compton radiate (off the CMB) at $\sim$GeV energies and synchrotron radiate at microwave frequencies.
However, given the severe radiative energy losses experienced by electrons, 
the hard spectrum, uniform intensity, vast extension, and energetics of the bubbles render the origin of this
particle population extremely mysterious \cite{Finkbeiner2004,Dobler2008,Dobler2009,McQuinn2010,Su2010}.
In particular, even accounting only for energy losses on the CMB, transport of $\geq$TeV, IC-radiating electrons to the requisite distances from the plane would require velocities of $> 0.03 \ c$, too fast for a Galactic wind (though an AGN jet potentially offers a suitable delivery mechanism in this connection \cite{Guo2011}).

A viable alternative to the idea that CR {\it electrons} generate the Bubble $\gamma$-rays is, however, that they originate in the collisions between cosmic ray {\it ions} and the Bubbles' low-density plasma.
In fact, the wind out of the GC we have identified above carries a power of $\sim 10^{39}$ erg/s out of the region, as we have emphasised, and this is precisely enough, in steady state and assuming the Bubbles represent thick targets to the injected protons, to sustain the observed 1-100 GeV luminosity of the Bubbles of $4 \times 10^{37}$ erg/s.
This scenario explains many aspects of the Bubbles' non-thermal phenomenology.
Firstly, the hard spectrum of the $\gamma$-rays is explained: in contrast to the situation in the Galactic disk where energy-dependent confinement implies a steepening of the in situ spectrum of CRs away from their injection distribution, the  Bubble CRs, trapped independently of energy by hypothesis, follow their injection distribution.
Likewise the  hard-spectrum `WMAP haze' \cite{Finkbeiner2004,Dobler2008}, coincident at lower Galactic latitudes with the $\sim$ GeV $\gamma$-ray emission, is explained in this scenario as a result of {\it secondary} electron synchrotron emission which, again, would provide a signal of precisely the right luminosity.
(We note in passing that decay of charged mesons -- which leads to secondary electrons and positrons -- also produces high-energy neutrinos and that -- if our scenario is correct -- the Bubbles should also be a significant source for a future, km$^3$ class, Northern Hemisphere neutrino telescope.)
Finally, a robustly-detected \cite{Su2010} down-turn in the Bubbles' SED below $\sim$GeV is explained as a natural result of the kinematics of neutral pion decay.

The `cost' of this scenario is the very long timescales implied: given the low-density of the Bubble plasma, $\lesssim 0.01$ cm$^{-3}$, the $pp$ loss time blows out to $\gtrsim 5$ Gyr and the Bubbles are required to have existed for at least this time (in order that steady state be satisfied) and to effectively trap CR protons (up to at least $\sim$TeV) over the same timescale.
Thus one requires that the GC has sustained injection of $\sim 10^{39}$ erg/s in CRs into the base of the Bubbles for multi Gyr timescales.
Such might seem hard to credit but is not unreasonable upon further reflection: the morphology of the Bubbles clearly privileges the GC and the GC is perhaps the single, spatially-localized site in the Galaxy where SF over multi-Gyr timescales is assured \cite{Serabyn1996,Figer2004}.
Note that while by no means required by our scenario, it is interesting that the {\it current} level of GC star-formation -- and resulting cosmic ray luminosity -- is close to the time-averaged value required in our scenario for the origin of the Bubbles.
This suggests a system in steady state.

A final couple of notes are in order. 
Firstly, aside from the sustained star-formation occurring in the GC, an a priori suspect to ultimately power the Bubble emission is the central, supermassive black hole (SMBH) \cite{Su2010,Guo2011}.
In the sort of hadronic scenario we have explored this would -- just as for star-formation and concomitant supernova activity -- be required to generate a time-averaged power of $\sim 10^{39}$ erg/s in CRs.
It is interesting in this context, then, 
that such a cosmic ray luminosity is rather close to the {\it minimum} required by analysis of the central, point-like
GeV $\gamma$-ray source under the assumption that it is hadronic in nature \cite{Chernyakova2011}.

Secondly, we note that Socrates et al. have recently postulated \cite{Socrates2008} the existence of an `Eddington limit in cosmic rays' following the logic that, as
cosmic rays diffuse outwards from some central source, they exchange momentum with surrounding gas via scattering on `frozen-in' magnetic field inhomogeneities and may, therefore, arrest gas accretion beyond some limiting luminosity. This is roughly estimated to be $L_{Edd}^{CR} \sim 10^{-6} L_{Edd}$.
For the case of the GC SMBH (of mass $\sim 4 \times 10^6 \msun$) or the central nuclear star cluster (of mass $\sim3 \times 10^7 \msun$: \cite{Launhardt2002}) the GC's {\it Central Massive Object} has a rough `cosmic ray Eddington limit'  in the range $(0.5-4) \times 10^{39}$ erg/s, again interestingly close to the value required to sustain the Bubbles' current $\gamma$-ray emission.
In conclusion, over the last few billions years, the GC seems to have emitted a time-averaged  cosmic ray luminosity of close to the maximally-allowed value consistent with sustained accretion.
We finally remark that, regardless of
whether it is GC star-formation or low-level but sustained activity of the SMBH that energises the Bubbles, in our hadronic scenario these remarkable structures
constitute a perfect calorimetric recording of Galactic centre
activity over the history of the Milky Way.

\section{Acknowledgments}

I gratefully acknowledge the contribution of my collaborators: David Jones, Felix Aharonian, Casey Law, Fulvio Melia, Tomo Oka, and J{\" u}rgen Ott.

\end{document}